\documentstyle{l-aa}
\begin{document}

\thesaurus{08.16.4, 08.15.1, 08.09.2 HS\,2324+3944}

\title{HS\,2324+3944: discovery of non-radial pulsations in a hydrogen-rich
PG\,1159 star
\thanks{Based on observations obtained at the Loiano Observatory, Italy.}}

\author{Roberto Silvotti}

\institute{Dipartimento di Astronomia, Universit\`a di Bologna, 
via Zamboni 33, I-40126 Bologna, Italy}

\offprints{silvotti@astbo3.bo.astro.it}

\date{Received February ..., accepted ......, 1996}

\maketitle

\begin{abstract}

HS\,2324+3944 is a peculiar PG\,1159 star, with a high amount of H in its
atmosphere (Dreizler et al. 1995).
Its location in the $\log T_{\rm eff}$ -- $\log g$ plane is well inside the
GW Vir instability strip.
In this paper I report the results of two photoelectric observations of
HS\,2324+39, which clearly show that the luminosity of this star presents
periodical variations with a period of (2140 $\pm$ 11) s.
This photometric behaviour is most easily explained by the presence of
non-radial oscillations.
Therefore HS\,2324+39 is a new member of the GW Vir group,
and is characterized by the longest pulsation period found among these stars.
Moreover HS\,2324+39 appears to be the first pulsating PG\,1159 star with a
high H abundance in its atmosphere (H/He = 2 by number, Dreizler et al. 1995).
Were the pulsation mechanism based on the C/O cyclic ionization (Starrfield
et al. 1984) at work, the H abundance should drop to zero sharply in the
driving regions.
Such a strong decrease of hydrogen looks quite unlikely; for this reason
the presence of pulsations in HS\,2324+39 seems to be a very interesting
phenomenon.


\end{abstract}

\section{Introduction}

HS\,2324+3944 was recognized analyzing the Hamburg Schmidt Survey
plates (Hagen et al. 1995, Wisotzki 1994).
It has been classified as a lgEH\,PG\,1159 (Dreizler et al. 1995), following
the scheme of Werner (1992).
Dreizler et al. (1995, hereafter DWHE95) have made a detailed analysis
of the spectral characteristics of HS\,2324+39; here is a brief summary
of what they have found.
HS\,2324+39 is a ``hybrid PG\,1159 star'', showing H Balmer absorption features
in its spectrum, and it is the only ``hydrogen PG\,1159'' not surrounded
by a planetary nebula.
The high H abundance (H/He = 2.0 $^{+0.5}_{-0.6}$ by number) is about 10 times
the upper limit found by Werner (1995) for PG\,1159--035\,!
The abundances of He, C and N (C/He = 0.3, N/He $<$ 0.002) are ``normal'',
whereas the O abundance (O/He = 0.006 $\pm$ 0.004) is quite low respect to the
``standard'' PG\,1159 stars (O/He = 0.13 for PG\,1159--035).
The effective temperature is equal to (130\,000 $\pm$ 10\,000) K;
the surface gravity, corresponding to log $g$ = 6.2 $\pm$ 0.2, is one of the
lowest for the PG\,1159 stars without nebula.
In fact, HS\,2324+39 is located in a region of the
$\log T_{\rm eff}$ -- $\log g$ plane where most objects are central stars of
planetary nebulae (CSPN) (DWHE95 Fig. 5).

With these values for temperature and gravity, HS\,2324+39 results to be
well within the GW Vir instability strip.
Presently the group of the GW Vir stars is formed by 13 objects
\footnote{We refer here to the``pulsational definition'' of the GW Vir stars;
from spectroscopy the GW Vir stars are only 8 (4 with nebula, 4 without), the
remaining 5 pulsating CSPN are early WR stars.}
\hspace{-1.5mm}
(9 CSPN, 4 not showing a nebula),
whose luminosity has multi-frequency
variations, due to the presence of non-radial g-mode pulsations (Bradley 1995,
Silvotti et al. 1995 and references therein).
The typical values of the pulsation periods are 10--35 minutes for the variable
CSPN (also called PNNV), and 5--15 min for the pulsating stars without
nebula (also called DOV).

The driving mechanism proposed to explain the pulsations in all these
stars is the $\kappa-\gamma$ mechanism, based on the C/O cyclic ionization
(Starrfield et al. 1984).
It has been shown (Stanghellini et al. 1991) that even a small amount of H
in the driving layers can be sufficient to inhibit the pulsation mechanism.
Such a high sensitivity to the chemical composition could in principle explain
why not all the PG\,1159 stars do pulsate.

\begin{figure*}[ht]
\vspace{88mm}
\includegraphics{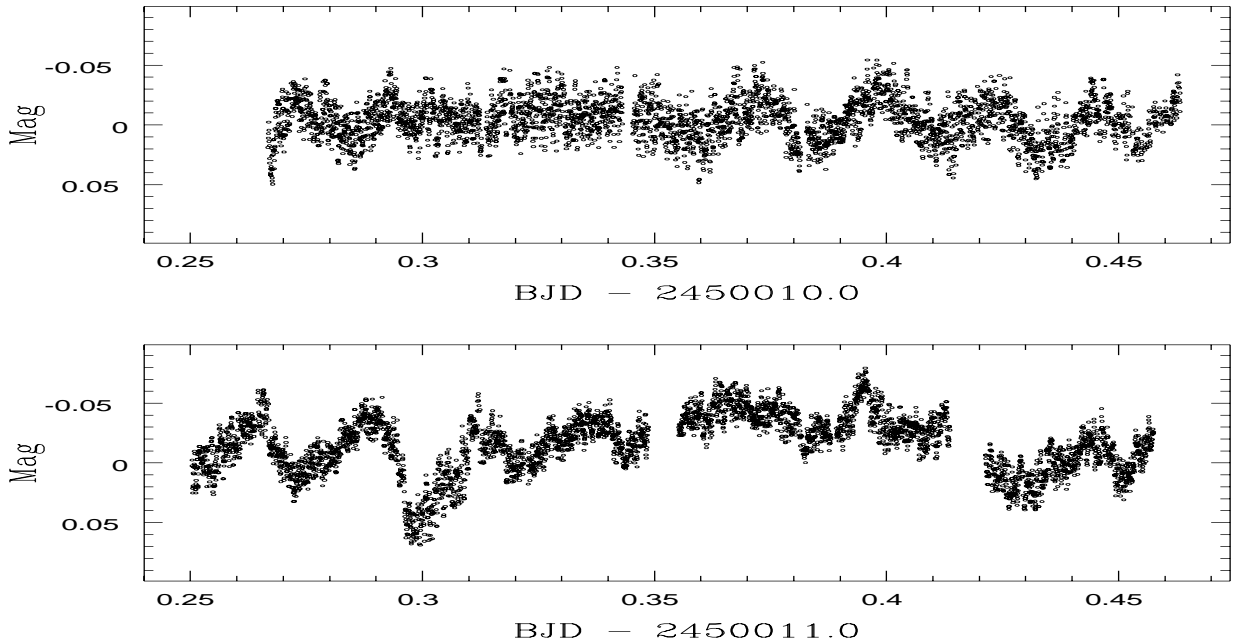}
\caption[]{Light curves of HS\,2324+3944 in October 19 and 20, 1995}
\end{figure*}

Now the high H abundance of HS\,2324+39 would suggest that no pulsations
are possible for this star.
Therefore the presence or the absence of pulsations in HS\,2324+39 seemed to be
an interesting test for the GW Vir pulsation mechanism.
For this reason I decided to investigate on the variability of HS\,2324+39.
The results of two photoelectric observations of this star are presented in
the following section of this paper.
A preliminary communication on the variability of HS\,2324+39 has been
already published (Silvotti 1995).

\section{Observations and results}

I observed HS\,2324+39 with the 2-head photoelectric photometer
(Piccioni et al. 1979, Bartolini et al. 1993) of the 1.5 m
Loiano telescope (Italy) on October 19 and 20, 1995, with no moon.
The tubes used, two EMI 9784 QB, have a maximum sensitivity in the B band.
Both observations were carried out without filter, with an integration
time of 2 s.
The comparison stars of the two observations were different (Silvotti 1995).
The light curves of the target and the comparison star for the two nights are
shown in Silvotti (1995).
Here I have selected the best curves, i.e. the difference of magnitude
between target and comparison star for the 19/10 observation, and HS\,2324+39
alone for the 20/10 observation.
The light curves are shown in Fig. 1, where each point represents a 4 s
integration (the data were binned bringing the effective integration time
from 2 to 4 s).
In both light curves we can see several arc features with a mean period
of about half an hour.
The quiescent zone in the light curve of October 19 (Fig. 1 from about 0.3 to
0.35 fractional Barycentric Julian Day (BJD)) could be caused by the
interference of different frequencies.

From the discrete Fourier transform of the best night (19/10) we obtain a
period of (2130 $\pm$ 70) s, with an amplitude of 10 mmag.
The amplitude spectrum is presented in Fig. 2, together with the spectrum of a
sinusoid with the same time distribution, same period and same amplitude
(spectral window).
If we make the discrete Fourier transform of both nights together
(32 $\%$ duty cycle), the main peak is at about 2130 s (amplitude 10 mmag),
in very good agreement with the values obtained on the first night.
Things are a bit different considering the second night alone:
the spectrum shows
two peaks at (2220 $\pm$ 80) and (1890 $\pm$ 50) s.
The amplitudes, which are respectively 14 and 12 mmag, are probably increased
by the sky noise
\footnote{The data of the second night were reduced using only the counts of
the target.
I adopted this procedure because the comparison star was
gone off the aperture for about one hour (Silvotti 1995).
In any case, even using the difference of magnitude between target and
comparison star, the light curve does not change significantly.
The reason is that the comparison star is about 2 magnitudes brighter than
HS\,2324+39.
Therefore channel 2 effectively compensates only for the transparency
variations, not the sky variations. 
The best solution to this problem is certainly to use a 3 channel photometer.
If only a 2 channel photometer is available, the best thing to do is probably
to choose a comparison star having about the same magnitude as the target,
as it was done in the second observation of HS\,2324+39.
In this manner both the transparency and the sky variations during the night
can be partially compensated for.}
\hspace{-1.5mm}.
It is not clear whether these two periodicities are real or not.
The higher period (2220 s $\pm$ 80) is not incompatible with the value found
on the first night (2130 s $\pm$ 70).
\begin{figure}[ht]
\vspace{60mm}
\includegraphics{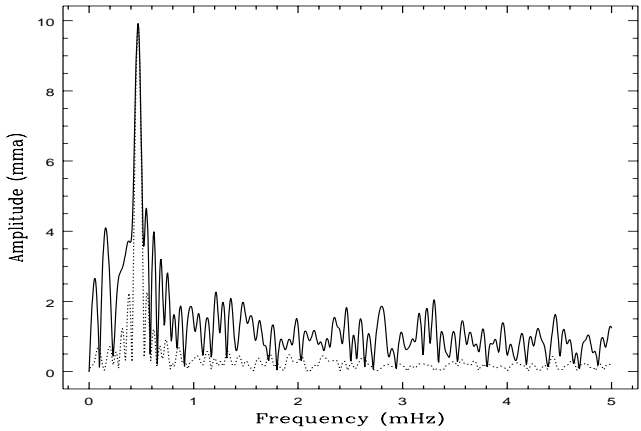}
\caption[]{Amplitude spectrum (solid line) and spectral window (dot line)
 of HS\,2324+3944 on October 19, 1995}
\end{figure}
In any case it is easy to verify that the data of the two successive nights
are in phase, and can be very well folded over a single frequency (Fig. 3).
In this hypothesis, namely that the luminosity variations of HS\,2324+39 are
due to a single pulsation frequency, we obtain the following best pulsation
period $P$:
\begin{center}
$P = (2140.5 \pm 11.0) \, s$
\end{center}
The error has been estimated considering:
\[ \hspace{18mm} \frac{\Delta P}{P} = \frac{1}{4} \times \frac{1}{number\,of\,
phased\,cycles} \]
\noindent
The times $t_{min}$ of the minima are given by:
\begin{center}
$t_{min} = [2450010.28540 \pm 0.00100 + n \, (0.02477 \pm 0.00013)] \, BJD$
\end{center}
From Fig. 3 (upper part) the average shape of the pulse appears to be slightly
asymmetric, with the descent steeper than the climb.

\begin{figure}[ht]
\vspace{88mm}
\includegraphics{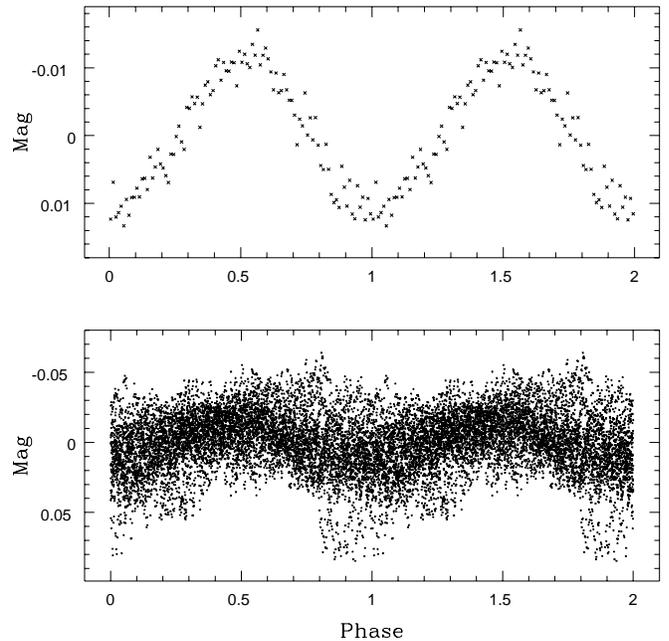}
\caption[]{The two observations of HS\,2324+3944 folded over a period of
2140.5 s. The top panel is obtained dividing the phase in 100 equal intervals
and taking the mean magnitude}
\end{figure}

\section{Summary and discussion}

HS\,2324+39 is a new H-rich peculiar PG\,1159 pulsating star.
It seems to be a very interesting object for several reasons:
\begin{enumerate}
\item{it is, so far, the only PG\,1159 pulsating star with a high H abundance
in its atmosphere (He/H=0.5 by number, DWHE95). This fact seems to be in
contrast with the C/O pulsation mechanism proposed by Starrfield et al. (1984),
unless we admit that the H abundance drops to zero sharply in the driving
regions. The hydrogen could be maintained in a thin surface layer by the
gravitational settling. This hypothesis would imply some limits to the strength
of the stellar winds and to the extension of the surface convection zone.
A weak stellar wind is also suggested by the shape of the blue part of the
C\,IV components near 4650 \AA \hspace{0.1mm} in the spectrum of HS\,2324+39
(DWHE95 Fig. 4, Leuenhagen 1995).

The edges of the theoretical GW Vir instability strip obtained by Starrfield
et al. (1984) and by Stanghellini et al. (1991) are calculated using the
Los Alamos opacities.
Starrfield et al. (1984) considered only carbon and oxygen in the surface,
whereas Stanghellini et al. (1991) contemplated also the presence of helium.
The presence of hydrogen was never considered.
The most realistic models, with 50 $\%$ carbon and 50 $\%$ helium by mass
(Stanghellini et al. 1991), are pulsationally unstable at effective
temperatures much lower than the real GW Vir stars.
More recently Saio (1996), using the OPAL opacities, has obtained overstable
modes in models with a surface composition closer to the real GW Vir stars
(Y=0.38, X$_{\rm C}$=0.4, X$_{\rm O}$=0.2, Z=0.02). 
A model sequence with a 3 $\%$ (by mass) abundance of hydrogen was also 
computed; the stability of g-modes results to be hardly affected by the
existence of hydrogen.
The chemical abundances obtained by Werner (1995) for PG\,1159--035
(X$<$0.015, Y=0.32, X$_{\rm C}$=0.48, X$_{\rm O}$=0.165, taking Z $\simeq$ 0.02)
are quite similar to those used by Saio (1996).
But for HS\,2324+39 (X=0.20, Y=0.405, X$_{\rm C}$=0.365, X$_{\rm O}$=0.01,
DWHE95 considering Z=0.02) things are quite different: the H and O abundances,
compared with PG\,1159--035, are in practice exchanged.
Nevertheless, the location of HS\,2324+39 in the
$\log T_{\rm eff}$ -- $\log \Pi$
diagram of Saio (1996, Fig. 5) is consistent with models having stellar masses
between 0.58 and 0.60 $M_{\odot}$, in good agreement with the value of
0.59 $M_{\odot}$, obtained by DWHE95 comparing the location of HS\,2324+39
in the $\log T_{\rm eff}$ -- $\log g$ plane with evolutionary tracks.}
\item{The pulsation period of (2140 $\pm$ 11) s is the longest ever observed
in a PG\,1159 star.}
\item{The duration of the pulsation period and the position of HS\,2324+39
in the $\log T_{\rm eff}$ -- $\log g$ plane (DWHE95) would suggest the presence
of a planetary nebula (PN) around the star. Moreover all the other H-rich
PG\,1159 stars are CSPN.
Presently it seems that HS\,2324+39 does not have any PN remnant.
A recent observation made at Calar Alto (Werner 1996) confirms this thesis.}
\end{enumerate}

For all these reasons, a detailed study of the HS\,2324+39 pulsation should be
undertaken. With precision asteroseismology some open questions regarding the
structure and the mass of the external layers of HS\,2324+39 could be solved.
The best way to achieve this purpose would be to observe HS\,2324+39 with
the Whole Earth Telescope (Nather et al. 1990).
The peculiar characteristics of HS\,2324+39 make this star very interesting
and give us a possibility to improve our knowledge on the GW Vir pulsation
phenomenon.

\begin{acknowledgements}

I am pleased to thank Klaus Werner, who suggested to observe this star
and provided me with the finding chart.
I also would like to thank Gerald Handler for having made available the
preliminary results of his observations on HS\,2324+39, taken at the McDonald
Observatory (Texas), which were consistent with my results, and Stefan
Dreizler for interesting comments about the pulsation of the PG\,1159 stars.
Thanks also to Gianni Tessicini and Roberto Gualandi for the technical support
at the telescope.
Finally thanks to Darragh O'Donoghue for his visit to the Loiano Observatory
in the same period when the observations reported in this paper were carried
out.
This research was partially supported by the Italian ``Ministero per 
l'Universit\`a e la Ricerca Scientifica e Tecnologica'' (MURST).

\end{acknowledgements}

\end{document}